\documentclass[journal]{IEEEtran}
\usepackage{cite}
\ifCLASSINFOpdf
   \usepackage[pdftex]{graphicx}
\else
   \usepackage[dvips]{graphicx}
\fi
\usepackage{amsmath}
\ifCLASSOPTIONcompsoc
  \usepackage[caption=false,font=normalsize,labelfont=sf,textfont=sf]{subfig}
\else
  \usepackage[caption=false,font=footnotesize]{subfig}
\fi
 \let\MYorigsubfloat\subfloat
 \renewcommand{\subfloat}[2][\relax]{\MYorigsubfloat[]{#2}}
\usepackage{bm}
\usepackage{multirow}
\allowdisplaybreaks

\usepackage{lineno}
\usepackage{booktabs,multirow,amsthm,amsmath,mathrsfs}
\usepackage{algorithm}
\usepackage{algorithmic} 
\usepackage{amssymb} 

\usepackage{subfig}
\usepackage{setspace}
\usepackage{changepage}

\usepackage{color}
\begin{document}

%\title{{A Polynomial Time Approximation Scheme for Shortest-Longest Path Problem}}
\title{{An FPTAS for Shortest-Longest Path Problem}}
%\title{{Computing Bandwidth-Delay Optimal Paths via Segment Routing}}


%\author{Author~1}

\author{Jianwei~Zhang
	\thanks{Jianwei Zhang (janyway@outlook.com) is with the 
		School of Information Science and Engineering, 
		Yunnan University, Kunming, 650500, China.}% <-this % stops a space
}


\markboth{Technical Report, 2025}%
{Shell \MakeLowercase{\textit{et al.}}: Bare Demo of IEEEtran.cls for IEEE Journals}



\maketitle

\begin{abstract}
Motivated by multi-domain service function chain (SFC) orchestration, we define the shortest-longest path (SLP) problem, prove its hardness, and design an efficient fully polynomial time approximation scheme (FPTAS) using the dynamic programming (DP) and scaling and rounding (SR) techniques to compute an approximation solution with provable performance guarantee.
The SLP problem and its solution algorithm have theoretical significance in multicriteria optimization and also have application potential in QoS routing and multi-domain network resource allocation scenarios.
\end{abstract}


\IEEEpeerreviewmaketitle

\section{Introduction}
Suppose that we want to deploy \textit{service function chain} (SFC) across multiple routing domains using \textit{network function virtualization} (NFV). 
The total computing resources offered by each domain and accumulated along the path must be sufficient to host all the \textit{virtual network functions} (VNFs) of the SFC \cite{full_mesh}. 
Furthermore, the total cost (or delay, IGP weight) along the path is expected to be minimized.
With the assistance of \textit{topology aggregation} (TA) \cite{ta_sfc}, networking and computing resource information in a domain can be abstracted into virtual edges among border routers \cite{full_mesh}.
Then, a natural question is how to find a path simultaneously satisfies the cost constraint and the resource constraint over the abstracted virtual network.

Existing studies \cite{huang2011,xue2007,xue2008,improved2017} related to multi-constrained \textit{quality-of-service} (QoS) routing mainly focus on the case that 
all the metrics belong to the \textit{minimization type}, i.e., 
the metric sum along a path cannot be larger than some upper bound.
However, the aforementioned resource metric belongs to the \textit{maximization type}, i.e., the metric sum along a path cannot be smaller than some lower bound.
Although both of the two types are \textit{additive} metrics, they cannot be solved using existing approaches.
This is the theoretical motivation of this paper.

In this paper, we define the \textit{shortest-longest path} (SLP) problem, prove its hardness, and design an efficient \textit{fully polynomial time approximation scheme} (FPTAS) using the \textit{dynamic programming} (DP) and \textit{scaling and rounding} (SR) techniques to compute an approximation solution with provable performance guarantee.
The SLP problem and its solution algorithm have theoretical significance in multicriteria optimization and also have application potential in QoS routing and multi-domain network resource allocation scenarios.

\section{Model}
The network is modeled as a directed graph $G=(V, E, w_S, w_L)$, where $V$ represents the vertex set and $E$ the edge set.
The number of vertices and edges are denoted by $n$ and $m$, respectively.
Each edge $e$ is associated with a performance metric pair $\left(w_{S}(e),w_{L}(e)\right)$.
The $S$-metric can be cost, delay, IGP weight, and etc.
The $L$-metric can be computing (or other types of) resource.
%Both of them are additive metrics.
%
The request $r=(s, t, W_S, W_L)$ is a unicast from $s$ to $t$ whose routing path $p$ satisfies ${w_S}(p) \le {W_S}$ and ${w_L}(p) \ge {W_L}$.

\textbf{Definition 1.} \textit{Feasible path}:
The path $p$ from $s$ to $t$ in $G$ that satisfies ${w_S}(p) \le {W_S}$ and ${w_L}(p) \ge {W_L}$ is a feasible path.

\begin{figure}[!h]
	\centering
	\vspace{-0.25cm}
	\includegraphics[angle=0, width=0.29\textwidth]{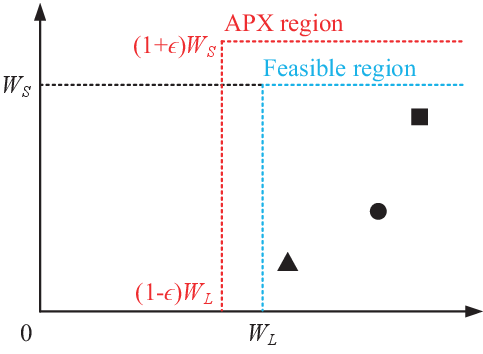}
	\caption{Feasible region and approximation region.}
	\label{fg:slp:space}
\end{figure}

\textbf{Definition 2.} \textit{Approximately feasible path}:
The path $p$ from $s$ to $t$ in $G$ that satisfies ${w_S}(p) \le {\left( 1+\epsilon  \right)W_S}$ and ${w_L}(p) \ge {\left( 1-\epsilon  \right)W_L}$ is an approximately feasible path. 
The factor $\epsilon \in (0,1]$ controls the approximation precision.
The feasible region and the \textit{approximation} (APX) region are illustrated by Fig.~\ref{fg:slp:space}.

\textbf{Definition 3.} \textit{Shortest-longest path (SLP) problem}:
Given a graph $G=(V, E, w_S, w_L)$ and a request $r=(s, t, W_S, W_L)$, 
the SLP problem is to compute a feasible path $p$ for $r$ in $G$.

\textbf{Theorem 1.}
The SLP problem is NP-complete.

\textit{Proof:}
First, whether a path is feasible for the SLP problem can be easily verifiable in polynomial time by definitions 1 and 3, implying that it is an NP problem.
Second, the problem includes the \textit{decision version} of the \textit{longest path problem}, which is NP-complete, as a special case, implying that it is at least an NP-complete problem.
Therefore, The SLP problem is NP-complete.
\qed

\begin{figure}[!h]
	\centering
	\includegraphics[angle=0, width=0.49\textwidth]{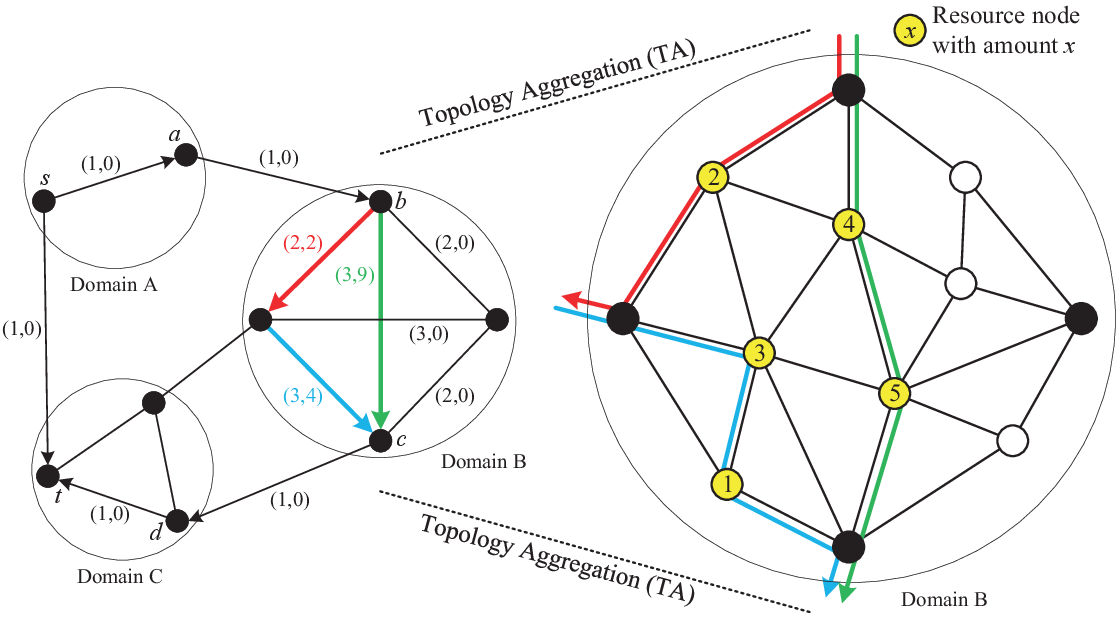}
	\caption{Motivation example. Only TA for domain B is shown.}
	\label{fg:slp:ta}
\end{figure}

Fig.~\ref{fg:slp:ta} is a motivation example regarding multi-domain network resource allocation.
The left part is an abstract multi-domain network after TA.
The right part is the TA details for domain B; other domains are omitted for clarity.
Assume that inter-domain links, i.e., links between border nodes, have unit cost metrics and zero resource metrics.
It is easy to verify that the only feasible solution for request $(s,t,9,9)$ is the path $s-a-b-c-d-t$.
The direct link (path) $s-t$ between domains A and C has zero resource and therefore cannot be a solution.
Note that our problem is based on the abstract network after TA.
The TA strategy itself permits great design flexibility and is beyond the scope of this paper.

\section{Algorithm}
Considering the NP-completeness of the SLP problem, we propose an approximation algorithm FPTAS-SLP, shown in Algorithm \ref{alg_slp}.
FPTAS-SLP includes four main steps \cite{huang2011,xue2007,xue2008}.
Step 1 is to scale and round the edge weights in the original graph using the approximation factor $\epsilon$.
Step 2 involves auxiliary graph construction.
Step 3 computes a shortest path w.r.t. the $S$-metric while executing \textit{edge repetition detection} (ERD) on the original graph.
Step 4 checks the feasibility.
\begin{algorithm}[!h]
	\caption{FPTAS-SLP}\label{alg_slp}
	\begin{algorithmic}[1]
		\REQUIRE $G$; $r$; $\epsilon$.
		\ENSURE An approximately feasible path ${p}^{\tau }$ for $r$ in $G$ with approximation ratio $(1 + \epsilon ,1 - \epsilon)$, if there exists one.
		\STATE Set the new weights $w_{k}^{\tau }(e)=\left\lceil \frac{{{w}_{k}}(e)}{{{W}_{k}}}\cdot \tau  \right\rceil, k=S,L$ and new bounds $W_{k}^{\tau }=\left\lceil \tau  \right\rceil, k=S,L$, where $\tau :=\frac{m}{\epsilon }$. 
		\STATE Construct an auxiliary graph ${{G}^{\tau }}=\left\{ {{V}^{\tau }},{{E}^{\tau }} \right\}$.
		The vertex set is ${{V}^{\tau }}=V\times \left\{ 0,1,...,\left\lceil \tau  \right\rceil  \right\}$.
		The edge set ${{E}^{\tau }}$ contains directed edges from vertex $(u,x)$ to $(v,y)$ such that $y=x+\min \left\{ w_{L}^{\tau }(u,v),\left\lceil \tau  \right\rceil -x \right\}$. The weight of all such edges is $w_{S}^{\tau }(u,v)$.
		\STATE Calculate the shortest path ${{p}^{\tau }}$ in ${{G}^{\tau }}$ from $(s,0)$ to $(t,\left\lceil \tau  \right\rceil)$, while ERD is performed in $G$.
		\STATE If $w_{S}^{\tau }({{p}^{\tau }})\le W_{S}^{\tau }$, output ${{p}^{\tau }}$; Else, output no solution.
	\end{algorithmic}
\end{algorithm}

\begin{figure}[!h]
	\centering
	\includegraphics[angle=0, width=0.47\textwidth]{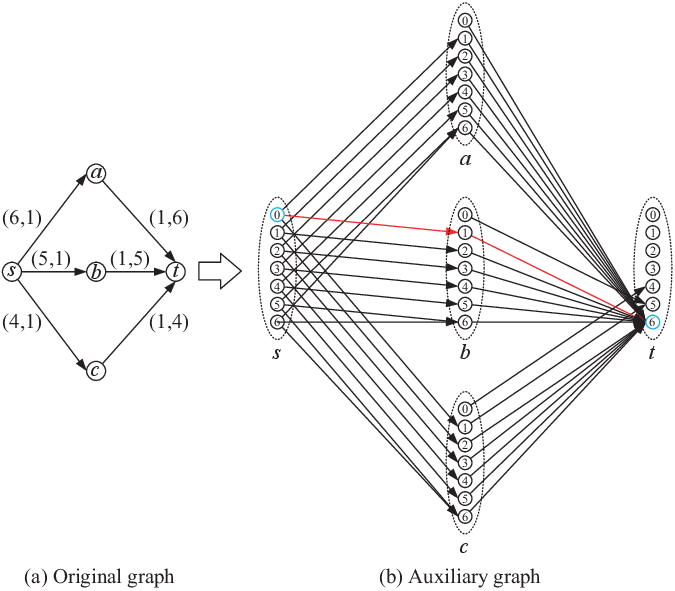}
	\caption{Illustrative example.
		$r=(s, t, 6, 6)$. $\epsilon=1$.}
	\label{fg:slp:agc}
\end{figure}

Fig.~\ref{fg:slp:agc} illustrates how to construct an auxiliary graph.
The auxiliary graph construction is a DP process.
The key differences between the auxiliary graph and the one in \cite{huang2011} are two-fold.
First, the edges between adjacent sub-vertices in a vertex are removed. For instance, there are no edges between $(s,0)$ and $(s,1)$.
Second, some edges are newly added. For instance, there is an edge between $(s,6)$ and $(c,6)$.
We note that the auxiliary graph \textit{may not be} a \textit{directed acyclic graph} (DAG).
It is worthwhile noticing an important feature of the auxiliary graph that 
any path reaching the last vertex is feasible w.r.t. the scaled and rounded $L$-metric bound, but not necessarily the $S$-metric bound.
For example, both of the paths $s \rightarrow  a \rightarrow t$ and $s \rightarrow b \rightarrow t$ can reach the last vertex $(t,6)$ and therefore satisfy the $L$-metric constraint,
while the former one violates the $S$-metric constraint and the latter one is the only feasible path in this example.

\begin{figure}[!h]
	\centering
	\includegraphics[angle=0, width=0.33\textwidth]{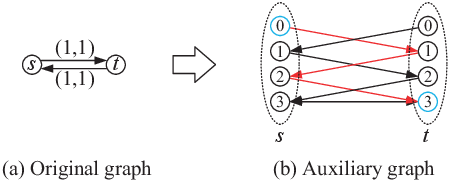}
	\caption{The necessity of ERD. $r=(s, t, 3, 3)$, $\epsilon=2/3$.}
	\label{fg:slp:loop}
\end{figure}

Fig.~\ref{fg:slp:loop} illustrates the necessity of ERD in Algorithm \ref{alg_slp}.
If the $L$-metric is interpreted as some type of resource, ERD guarantees that the resource can be utilized at most once.
Specifically, the path in red color does not cause edge repetition in the auxiliary graph while it does in the original graph.
Thus, it will be pruned from the solution space in the shortest path computation module (Step 3) of Algorithm \ref{alg_slp}.
In this sense, there does not exist a feasible solution under the setting $r=(s, t, 3, 3)$ and $\epsilon=2/3$.

\textbf{Theorem 2.}
The proposed FPTAS-SLP is guaranteed to compute a $\left( {1 + \epsilon ,1 - \epsilon} \right)$-approximation solution for the SLP problem, where $\epsilon \in (0,1]$, suppose that it is feasible.
%The time complexity is $O\left(\frac{n}{\epsilon}(m+n\log{n})\right)$. % Version 1: Without loop detection
The time complexity is $O\left(\tau n \log{\tau n}+\tau m^2 \right)$, where $\tau :=\frac{m}{\epsilon }$. % Version 2: With loop detection

\textit{Proof:}
We first analyze the time complexity \cite{huang2011,xue2007,xue2008}.
% Version 1: Without loop detection
%Since ${G}^{\tau }$ has $\tau n$ vertices and $O(\tau m)$ edges,
%the shortest path computation using Dijkstra's algorithm (optimized by Fibonacci heap) leads to the worst-case time complexity of 
%$O\left(\tau(m+n\log{n})\right)=O\left(\frac{n}{\epsilon}(m+n\log{n})\right)$.
%
%During each iteration, this only results in linear time complexity and has no impact on the overall time complexity.
%
% Version 2: With ERD
${G}^{\tau }$ has $\tau n$ vertices and $O(\tau m)$ edges.
For both ${G}$ and ${G}^{\tau }$, each node has $\theta = m/n$ downstream neighbors on average.
To realize the loop detection function, we only need to make a slight modification to Dijkstra's algorithm.
Specifically, each node records the current best path and also maintains the node mapping relationship between ${G}$ and ${G}^{\tau }$.
Dijkstra's algorithm on ${G}^{\tau }$ consists of $O(\tau n)$ outer loops. % and each outer loop .
Each outer loop contains one minimum extraction operation and $\theta$ relax operations (inner loops). % Assume min-extraction operation contains delete operation.
Using the Fibonacci heap, the minimum extraction operation takes $O(\log{\tau n})$.
In each relax operation, it takes $O(m)$ to perform ERD when using an array. 
Thus, the overall time complexity is $O\left(\tau n (\log{\tau n}+\theta m)\right)$,
i.e., $O\left(\tau n \log{\tau n}+\tau m^2 \right)$.
In the following, we prove the approximation performance of FPTAS-SLP.
The proof gains some ideas from \cite{huang2011,xue2007,xue2008}.

We denote by ${{p}^{\tau }}$ the optimal path returned by FPTAS-SLP.
Its feasibility w.r.t. the $L$-metric in the auxiliary graph yields its lower bound
\begin{equation}\label{eq:L:lb}\small
w_L^\tau ({p^\tau }): = \sum\limits_{e \in {p^\tau }} {\left\lceil {\frac{{{w_L}(e)}}{{{W_L}}} \tau } \right\rceil }  \ge \left\lceil \tau  \right\rceil  \ge \tau.
\end{equation}

Considering that the path length is at most $m$, the upper bound of $w_L^\tau ({p^\tau })$ is derived as
\begin{equation}\label{eq:L:ub}\small
\sum\limits_{e \in {p^\tau }} {\left\lceil {\frac{{{w_L}(e)}}{{{W_L}}}\tau } \right\rceil }  \le \sum\limits_{e \in {p^\tau }} {\left( {\frac{{{w_L}(e)}}{{{W_L}}}\tau {\rm{ + 1}}} \right)}  \le \frac{\tau }{{{W_L}}}{w_L}({p^\tau }) + m.
\end{equation}

Combining (\ref{eq:L:lb}) and (\ref{eq:L:ub}), we have
\begin{equation}\label{eq:L:1}\small
\tau  \le \frac{\tau }{{{W_L}}}{w_L}({p^\tau }) + m.
\end{equation}

Plugging into $\tau: =\frac{m}{\epsilon }$, some calculation on (\ref{eq:L:1}) yields
\begin{equation}\label{eq:L}\small
{w_L}({p^\tau }) \ge \left( {1 - \epsilon } \right){W_L}.
\end{equation}

The feasibility of ${{p}^{\tau }}$ w.r.t. the $S$-metric in the auxiliary graph yields its upper bound
\begin{equation}\label{eq:S:ub}\small
w_S^\tau ({p^\tau }): = \sum\limits_{e \in {p^\tau }} {\left\lceil {\frac{{{w_S}(e)}}{{{W_S}}}\tau } \right\rceil }  \le \left\lceil \tau  \right\rceil  \le \tau  + 1 \le \tau  + m.
\end{equation}

Similarly, the lower bound of $w_S^\tau ({p^\tau })$ is derived as
\begin{equation}\label{eq:S:lb}\small
\sum\limits_{e \in {p^\tau }} {\left\lceil {\frac{{{w_S}(e)}}{{{W_S}}}\tau } \right\rceil }  \ge \sum\limits_{e \in {p^\tau }} {\left( {\frac{{{w_S}(e)}}{{{W_S}}}\tau } \right)}  = \frac{\tau }{{{W_S}}}{w_S}({p^\tau }).
\end{equation}

Combining (\ref{eq:S:ub}) and (\ref{eq:S:lb}), we have
\begin{equation}\label{eq:S:1}\small
\frac{\tau }{{{W_S}}}{w_S}({p^\tau }) \le \tau  + m.
\end{equation}

Plugging into $\tau: =\frac{m}{\epsilon }$, some calculation on (\ref{eq:S:1}) yields
\begin{equation}\label{eq:S}\small
{w_S}({p^\tau }) \le \left( {1 + \epsilon } \right){W_S}.
\end{equation}

To sum up, Eqs. (\ref{eq:L}) and (\ref{eq:S}) together complete the proof.
\qed

%From the opposite perspective, the theorem also implies that the SLP problem is \textit{strictly} infeasible if FPTAS-SLP cannot compute an approximation solution.

\section{Conclusion}
In this paper, we for the first time define the SLP problem and design an approximation algorithm FPTAS-SLP with provable performance guarantee.
There are several issues to be carefully considered in the future.
First, the solution returned by FPTAS-SLP is \textit{approximately feasible} rather than \textit{approximately optimal} just as in \cite{xue2007,xue2008}.
Second, as long as FPTAS-SLP outputs a feasible solution, it must be approximately feasible for the SLP problem, but not vice versa.
That is, it is theoretically possible that the SLP problem has a feasible solution while FPTAS-SLP outputs no feasible solution.
The direct cause is that 
it has not been proven that any feasible solution in the original graph is also feasible in the auxiliary graph due to the effect of scaling and rounding.
Another reason is that FPTAS-SLP integrates a simple ERD module into a standard shortest path algorithm to avoid loops, which may reduce the solution space.
\end{document}